\documentclass{article}

\PassOptionsToPackage{square,numbers,sort&compress}{natbib}

\usepackage{authblk}

\usepackage[utf8]{inputenc} 
\usepackage[T1]{fontenc}    
\usepackage[hidelinks]{hyperref}       
\usepackage{url}            
\usepackage{booktabs}       
\usepackage{amsfonts}       
\usepackage{bm}
\usepackage{nicefrac}       
\usepackage{microtype}      
\usepackage{xcolor}         
\usepackage{bbm}
\usepackage{dsfont}
\usepackage[
  separate-uncertainty = true,
  multi-part-units = repeat
]{siunitx}
\usepackage{fixmath}
\usepackage[top=2cm,bottom=2cm,left=3cm,right=2.5cm]{geometry}

\usepackage{graphicx}
\usepackage{amsmath,mathtools,amsthm}
\usepackage{enumitem}
\graphicspath{{figs/}}
\usepackage{wrapfig}
\usepackage{algorithm}
\usepackage{algpseudocode}
\usepackage{caption}
\usepackage{subcaption}
\usepackage{comment}
\usepackage{xr}

\makeatletter
\newcommand*{\addFileDependency}[1]{
  \typeout{(#1)}
  \@addtofilelist{#1}
  \IfFileExists{#1}{}{\typeout{No file #1.}}
}
\makeatother

\usepackage{booktabs}

\usepackage{mathtools}
\mathtoolsset{showonlyrefs}

\usepackage{bbm}

\def\va{{\bm{a}}}
\def\vb{{\bm{b}}}

\def\vq{{\bm{q}}}

\def\vs{{\bm{s}}}

\def\vv{{\bm{v}}}

\def\vx{{\bm{x}}}

\def\vz{{\bm{z}}}

\def\mA{{\bm{A}}}

\def\mI{{\bm{I}}}

\def\mW{{\bm{W}}}

\newcommand*\samethanks[1][\value{footnote}]{\footnotemark[#1]}

\title{Quantum Sparse Coding}

\author[1,2]{
Yaniv~Romano\thanks{Equal contribution.}}
\author[2]{Harel~Primack\samethanks}
\author[2]{Talya~Vaknin\samethanks}
\author[2]{Idan~Meirzada}
\author[2]{Ilan~Karpas}
\author[2]{Dov~Furman}
\author[2]{Chene~Tradonsky}
\author[2]{Ruti~Ben~Shlomi}

\affil[1]{Departments of Electrical and Computer Engineering and of Computer Science, Technion IIT, Israel}
\affil[2]{LightSolver LTD., Tel Aviv, Israel}

\begin{document}

\date{}

\maketitle

\begin{abstract}
The ultimate goal of any sparse coding method is to accurately recover from a few noisy linear measurements, an unknown sparse vector. Unfortunately, this estimation problem is NP-hard in general, and it is therefore always approached with an approximation method, such as lasso or orthogonal matching pursuit, thus trading off accuracy for less computational complexity. In this paper, we develop a quantum-inspired algorithm for sparse coding, with the premise that the emergence of quantum computers and Ising machines can potentially lead to more accurate estimations compared to classical approximation methods. To this end, we formulate the most general sparse coding problem as a quadratic unconstrained binary optimization (QUBO) task, which can be efficiently minimized using quantum technology.
To derive at a QUBO model that is also efficient in terms of the number of spins (space complexity), we separate our analysis into three different scenarios. These are defined by the number of bits required to express the underlying sparse vector: binary, 2-bit, and a general fixed-point representation. We conduct numerical experiments with simulated data on LightSolver's quantum-inspired digital platform to verify the correctness of our QUBO formulation and to demonstrate its advantage over baseline methods.
\end{abstract}


\baselineskip=\normalbaselineskip

\section{Introduction}\label{sec:intro}

A ubiquitous problem in machine learning, statistics, and signal processing is to accurately estimate an unknown sparse vector from a few noisy linear measurements. This estimation problem, which we refer to as \emph{sparse coding}, is at the heart of the field of compressed sensing, revealing that under sparsity assumptions it is possible to successfully recover a signal that sampled significantly below the Nyquist rate~\cite{candes2006stable,donoho2006compressed}. This, in turn, led to a dramatic increase in magnetic resonance imaging (MRI) scanning session speed~\cite{lustig2007sparse}. Another exciting application that also builds on the sparsity assumption is unsupervised representation learning, i.e., given high-dimensional input data, such as an image, finding a low-dimensional representation that captures the intrinsic underlying structure in the input~\cite{engan1999method,aharon2006k,mairal2009online}. These representations are often used in image restoration tasks to effectively remove noise (denoising)~\cite{elad2006image,dong2011sparsity}, fill-in missing pixels (inpainting)~\cite{shen2009image,fadili2009inpainting,sulam2016trainlets}, and to achieve high quality digital zoom (super-resolution) \cite{fadili2009inpainting,yang2010image,zeyde2010single,romano2014single}. Sparsity also plays a key role in linear regression when given a large pool of features, to form a predictive rule that estimates an unknown response using a smaller, \emph{interpretable} subset of features that manifests the strongest effects~\cite{tibshirani1996regression,chen1994basis,chen2001atomic,efron2004least}.

To formalize the sparse coding problem, which is central for tackling the aforementioned applications, we consider the following linear model:
\begin{equation}
    \vb = \mA\vx + \vv,
\end{equation}
where $\mA$ is a matrix of size ${M\times N}$, the vector $\vx$ is of length $N$, and $\vv$ is a noise vector of length~${M}$. In this paper, we focus on a challenging setting in which $M \ll N$, where a crucial assumption we make is that the vector $\vx$ is $k$-sparse, i.e., it contains only $k$ non-zero elements with $k \ll N$~\cite{donoho2006compressed,candes2006stable,elad2010sparse}. In the context of sparse linear regression, the vector $\vb$ is called a \emph{response}, the columns of $\mA$ are the \emph{features}, and the rows correspond to the \emph{observations}. 
In the context of unsupervised representation learning, the matrix $\mA$ is called a \emph{dictionary}, whose columns are referred to as basis elements. The assumption in this literature is that the underlying signal can be represented as a \emph{sparse} linear combination of a few columns taken from the dictionary $\mA$, providing in a form of a compression effect~\cite{engan1999method,aharon2006k,mairal2009online}.

Suppose we are given a matrix $\mA$ and a noisy vector $\vb$, and our goal is to find a sparse solution $\hat{\vx}$ to the following optimization problem:
\begin{align}
\label{eq:k-sparse-problem}
    \hat{\vx} = \arg \min_{\vx} \|\mA\vx - \vb\|_2^2 \  \ \text{subject to} \ \|\vx\|_0=k,
\end{align}
where $\|\vx\|_0$ is the $L_0$ pseudo-norm, counting the number of non-zeros in $\vx$. Unfortunately, the above is a non-convex optimization problem, and finding a sparse solution that minimizes~\eqref{eq:k-sparse-problem} is NP-hard in general~\cite{davis1997adaptive,natarajan1995sparse}: intuitively, to solve \eqref{eq:k-sparse-problem}, we should sweep over all possible combinations of the $k$-sparse vectors and choose the one that minimizes the squared error term. Problem \eqref{eq:k-sparse-problem} is often expressed in the following Lagrangian form:
\begin{align}
\label{eq:lambda-problem}
    \hat{\vx} = \arg \min_{\vx} \|\mA\vx - \vb\|_2^2 + \lambda \|\vx\|_0,
\end{align}
where a proper choice of $\lambda$---controlling the strength of the sparsity penalty (larger $\lambda$ leads to sparser $\hat{\vx}$)---may render \eqref{eq:k-sparse-problem} and \eqref{eq:lambda-problem} to yield the same solution. The desire to find a solution for the non-convex sparse coding problem has led to practical approximation algorithms, which are supported by theoretical guarantees. For example, the well-known lasso~\cite{tibshirani1996regression} and basis pursuit~\cite{chen2001atomic} algorithms suggest replacing the non-convex $L_0$ with the convex $L_1$ norm in \eqref{eq:lambda-problem}. This choice is attractive since one can utilize the convexity of the modified objective, and minimize it efficiently~\cite{daubechies2004iterative,beck2009fast,combettes2005signal}. Another popular algorithm is the orthogonal matching pursuit (OMP)~\cite{mallat1993matching,pati1993orthogonal,tropp2004greed}---a method that greedily picks one non-zero at a time in lieu of exhaustively sweeping over all the possible solutions. Loosely speaking, following \cite{elad2010sparse,tropp2006just}, these methods perform particularly well when (i) the sparsity level $k$ of the underlying $\vx$ is ``small enough''; and (ii) the columns of $\mA$ are of low coherence, implying that any subset of $k$ columns is almost an orthonormal submatrix~\cite{candes2005decoding}. In fact, under certain assumptions on $k$, the correlation structure of $\mA$, and the noise level, lasso and OMP can provably find the correct locations of the non-zero elements in the unknown $\vx$, and, in turn, obtain an estimate $\hat{\vx}$ that is close to $\vx$, e.g., in the Euclidean distance~\cite{tropp2004greed,tropp2006just,candes2006stable,donoho2006compressed,elad2010sparse}.

The NP-hardness nature of the sparse coding problem brings us to the fascinating and ever-evolving field of quantum computing. The rise of quantum or quantum-inspired technologies unlock new opportunities to solve more accurately NP-hard problems than classic computing machines. Notable examples include the traveling salesman problem~\cite{lucas2014ising}, protein folding~\cite{dill2012protein}, and graph clustering~\cite{schaeffer2007graph}. Many of these problems can be cast as a quadratic unconstrained binary optimization (QUBO) problem~\cite{patton2019efficiently}, a formulation that is of particular interest to us as well. Furthermore, there has been growing activity in leveraging quantum technologies to tackle machine learning problems, such as linear regression and classification \cite{arthur2021qubo}, unsupervised clustering \cite{arthur2021qubo,otterbach2017unsupervised}, and even for improving deep generative models \cite{gao2020high}.

\subsection{Our contribution}

Due to the computational difficulty of minimizing the original problem \eqref{eq:lambda-problem}, it is always being replaced by one of its surrogates, e.g., lasso and OMP. However, in many cases, the conditions for guaranteeing the validity of the surrogates are either unknown or not satisfied. This paper ``goes back to basics'' and introduces novel quantum machine learning algorithms, directly minimizing the non-convex problem~\eqref{eq:lambda-problem} with the ultimate goal of obtaining more accurate solutions than standard approximation algorithms. Concretely, we show how to formulate \eqref{eq:lambda-problem} as a QUBO problem, which, in turn, can be minimized efficiently using cutting-edge solving platforms, such as quantum computers, Ising machines~\cite{meirzada2022lightsolver}, or well-known heuristics including Simulated Annealing and Tabu Search~\cite{glover1989tabu,glover1990tabu2}. Our derivation, presented in Section~\ref{sec:quantum-coding}, covers (i) a simple case where each entry in the unknown sparse vector $\vx$ is binary (1-bit representation); (ii) a more flexible setting, using a 2-bit representation per element in $\vx$; and (iii) the most general formulation where $\vx$ is represented using any arbitrary fixed-point precision. We also analyze the complexity of the proposed formulation with respect to the number of qubits (or spins) required to solve the QUBO problem. Naturally, the binary and 2-bit cases are highly efficient in terms of space complexity---we show how to form the QUBO problem without introducing any auxiliary (ancilla) spins. By contrast, the most general formulation in which $\vx$ can have fractional, non-integer entries comes at the cost of increasing the space complexity. Yet, our formulation is highly efficient as it requires only one ancilla spin for each entry in $\vx$, as we carefully argue in Section~\ref{sec:qubo_l0}.
Numerical experiments, given in Section~\ref{sec:res}, demonstrate the superiority of our QUBO-based solutions over the classic and most commonly used approximation algorithms---lasso and OMP. Specifically, we minimize the proposed QUBO problem via the quantum-inspired computing framework of LightSolver~\cite{meirzada2022lightsolver}, and show how our method tends to produce more accurate estimations of the unknown $\vx$ in regimes where the sample size is small or when the number of non-zero elements to be estimated is relatively large.

\subsection{Related work}

The general QUBO problem can be written as~\cite{patton2019efficiently}
\begin{equation}
\label{eq:qubo}
    \min_{\vq \in \mathbb{B}^D} \vq^{\top} \mW \vq,
\end{equation}
where $\mathbb{B}=\{0,1\}$ a binary set and $\mW \in \mathbb{R}^{D \times D}$ is a real-valued QUBO matrix. The ability to efficiently minimize~\eqref{eq:qubo} using quantum computers paves the way to the design of powerful solvers for fundamental machine learning problems. Importantly, QUBO optimization becomes appealing when the runtime of quantum-based algorithms is smaller than that of classic computers. Motivated by this capability, \cite{jun2021qubo} presented a QUBO formulation for solving a linear system of equations $\mA\vx=\vb$. The authors of \cite{arthur2021qubo} broadened this scope even further and showed how to fit least squares regressors, support vector machine (SVM) classifiers, and balanced k‑means clustering algorithms, by translating these problems into a QUBO scheme, too. Notably, the above algorithms were shown to have improved time and space complexities compared to classic SVM and balanced k-means clustering algorithms, and similar complexity to that of classic least squares solvers. Our work enriches the quantum machine learning toolbox by introducing a novel QUBO formulation for NP-hard sparse coding problems. Stressing this point, since we express the sparsity penalty in general QUBO terms, we could augment it to the QUBO SVM method of \cite{arthur2021qubo} and tackle the sparse SVM problem \cite{bradley1998feature} instead. We believe such an extension could be of great interest by itself, and we leave it for future work.

Recently, \cite{wezeman2022quantum,ayanzadeh2019quantum,ayanzadeh2020ensemble} presented a QUBO formulation for solving a special case of \eqref{eq:lambda-problem}---also referred to as binary compressive sensing---for which the unknown $\vx$ is sparse and \emph{binary}. These quantum-based solutions were shown to yield improved performance compared to the classic lasso method, an observation that is also corroborated by our experiments. Our paper builds upon the above line of work and extends the QUBO formulation to \emph{non-binary} $\vx$, represented in fixed-point arithmetic. A biological problem that is tightly connected to binary sparse signal recovery is the estimation of the transcription factor in DNA binding. By making a linear modeling assumption, \cite{li2018quantum} expressed the gene sequence as a QUBO problem with \emph{binary} sparse regularization. The authors compared their quantum-based estimation to existing baseline methods (including lasso) and reported that the most significant improvement is obtained in small sample size regimes. This observation is also supported by our experiments, although conducted in a different context as well as in the challenging setting where $\vx$ is a non-binary vector. The method reported in \cite{ide2022sparse} is the closet to ours, as it presents a QUBO formulation for the recovery of sparse signals represented in fixed-point arithmetic. Our formulation, however, is much more efficient with respect to the number of spins required to derive the QUBO model. Let $P$ be the number of bits defining the resolution of the fixed point representation. To present~\eqref{eq:lambda-problem} as a QUBO problem for $P\geq3$, the approach presented in~\cite{ide2022sparse} introduces $P-2$ auxiliary spins per variable while ours uses only one!
We conclude with \cite{ferrari2022towards}, which ``translated to QUBO'' several feature selection methods. Instead of minimizing \eqref{eq:lambda-problem}, the authors of  \cite{ferrari2022towards} assigned each feature with an information-theoretic measure of importance---e.g., its correlation with the target variable---and selected the ones that deemed to be the most important. This approximation allows handling a non-binary $\vx$ by assigning a single binary spin per feature, indicating whether the feature should be selected or not, akin to binary compressive sensing problems. Importantly, the formulations of the feature selection methods in \cite{ferrari2022towards} are substantially different than ours---we aim to tackle the original, ubiquitous sparse coding problem \eqref{eq:lambda-problem} in its most general form, rather than an alternative formulation of it.

\section{A motivating example} \label{sec:motivation}

Before describing the proposed QUBO formulations, we pause to present a motivating, small-scale  example that demonstrates the advantage of an exhaustive method for solving the sparse coding problem compared to classical approximation methods. To this end, we generate a matrix $\mA$ with $L_2$-normalized columns, i.e., $\|\va_{i}\|_2=1$ for all $1\leq i\leq N$, where $\va_{i}$ is the $i$th column of $\mA$. We design the matrix $\mA$ such that its \emph{mutual coherence}, defined as~\cite{donoho2005stable,tropp2006just,elad2010sparse}
\begin{equation}
\label{eq:coherence}
    \mu(\mA) = \max_{i \neq j}|\va_i^{\top} \va_j|,
\end{equation}
is minimized.\footnote{A detailed procedure for generating matrices with low coherence is given in Appendix~\ref{app:genA}.} We make this design choice since low mutual coherence is crucial for the success of lasso and OMP in recovering the unknown $\vx$~\cite{elad2010sparse}. Put simply, we wish to show that an algorithm that exhaustively search for a solution for \eqref{eq:lambda-problem} is highly valuable, even when ``playing on the turf'' of lasso and OMP. 

\begin{figure}
     \centering
     \begin{subfigure}[b]{0.329\textwidth}
         \centering
         \includegraphics[width=\textwidth]{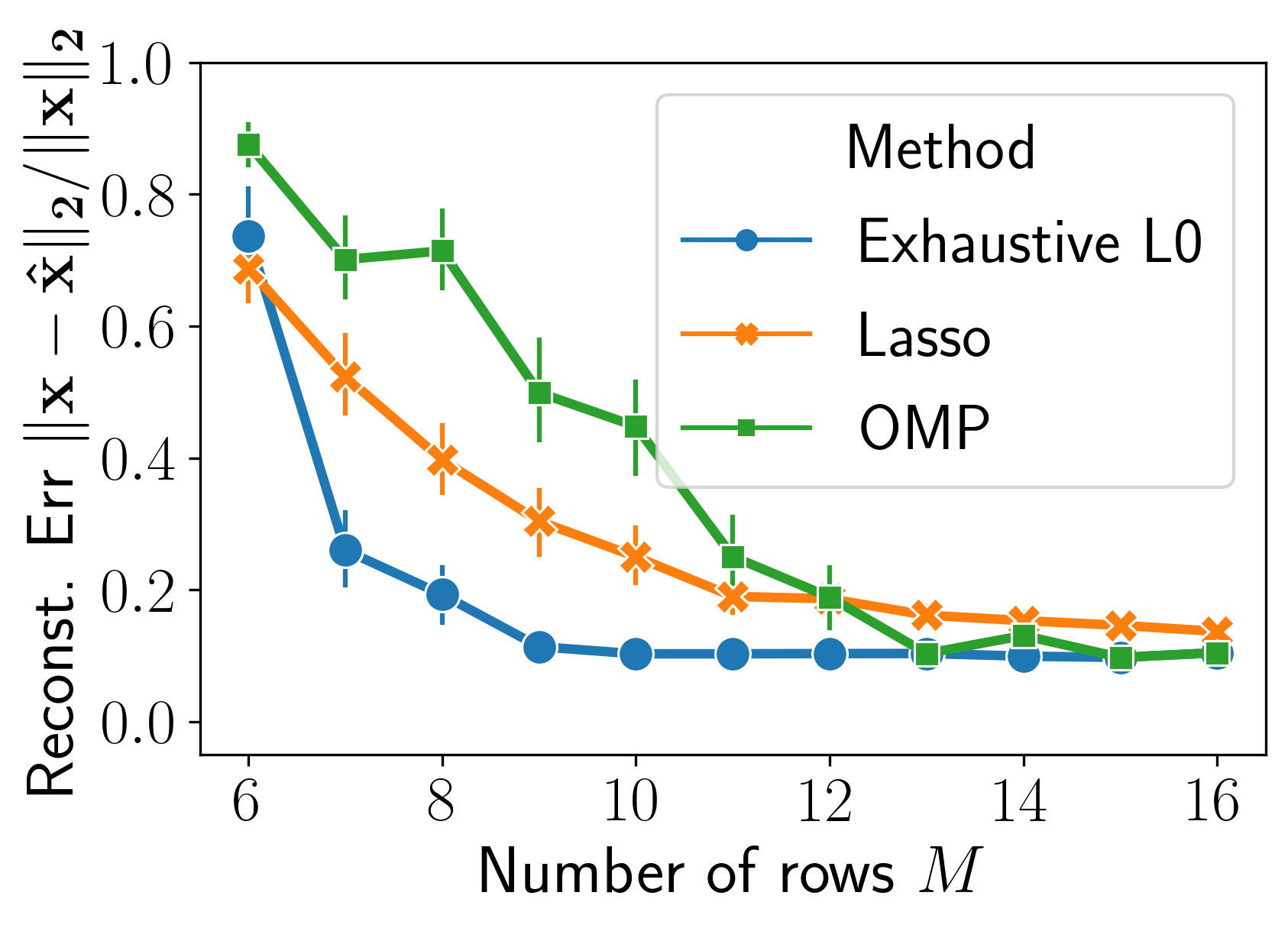}
         \includegraphics[width=\textwidth]{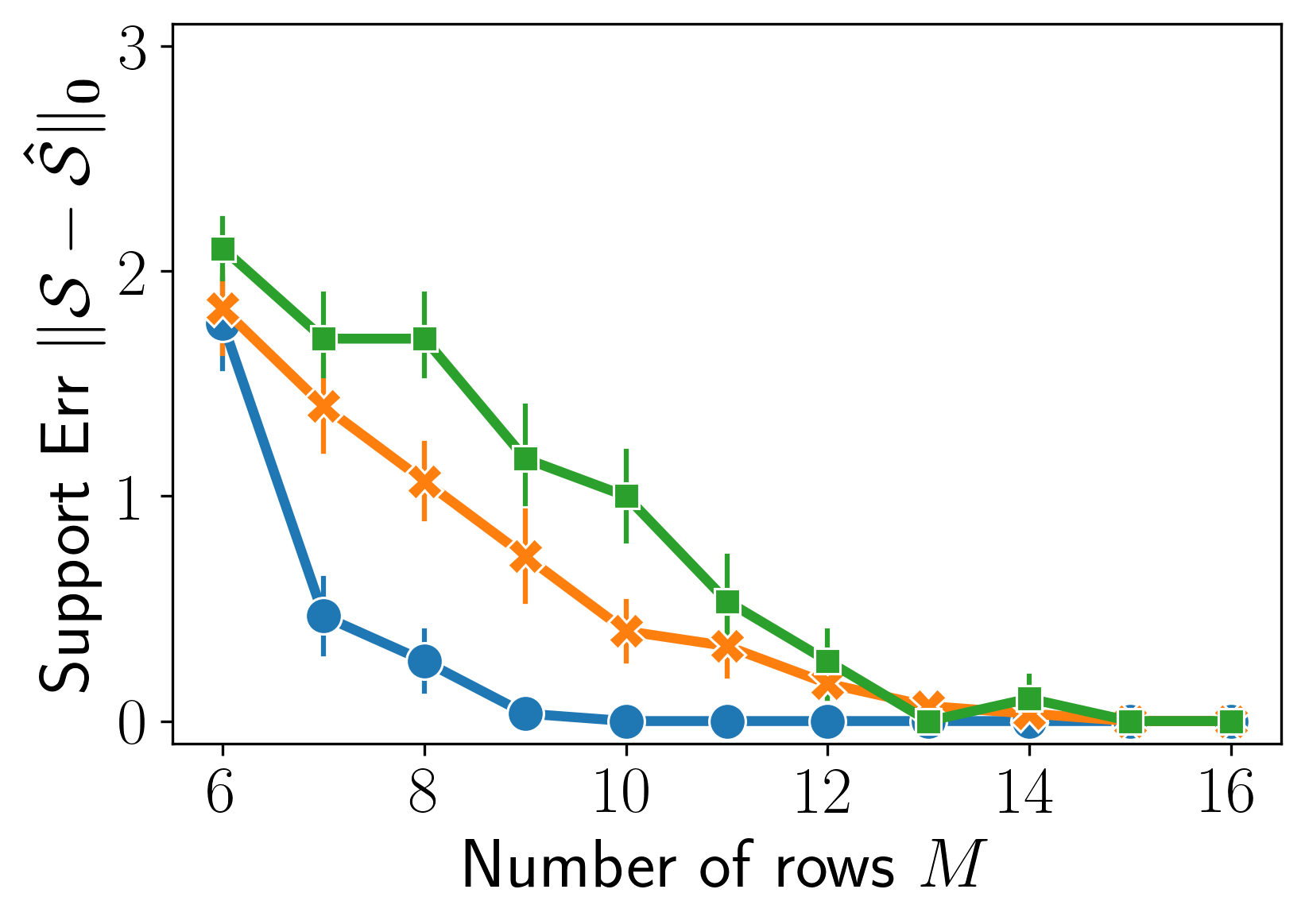}
         \caption{Cardinality $k=3$, noise $\sigma=0.1$.}
         \label{fig:mot_rec_vs_m}
     \end{subfigure}
     \hfill
     \begin{subfigure}[b]{0.329\textwidth}
         \centering
         \includegraphics[width=\textwidth]{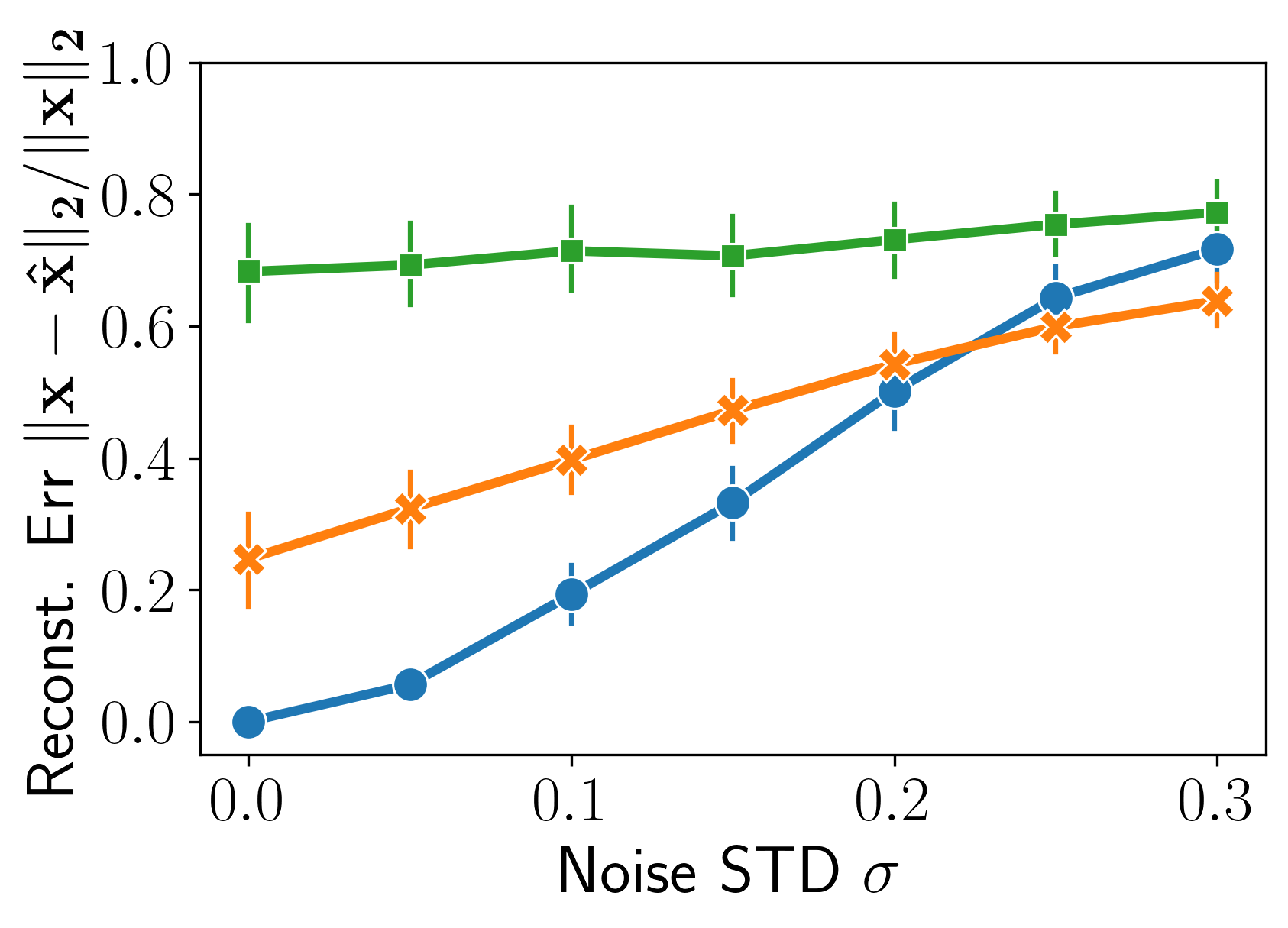}
         \includegraphics[width=\textwidth]{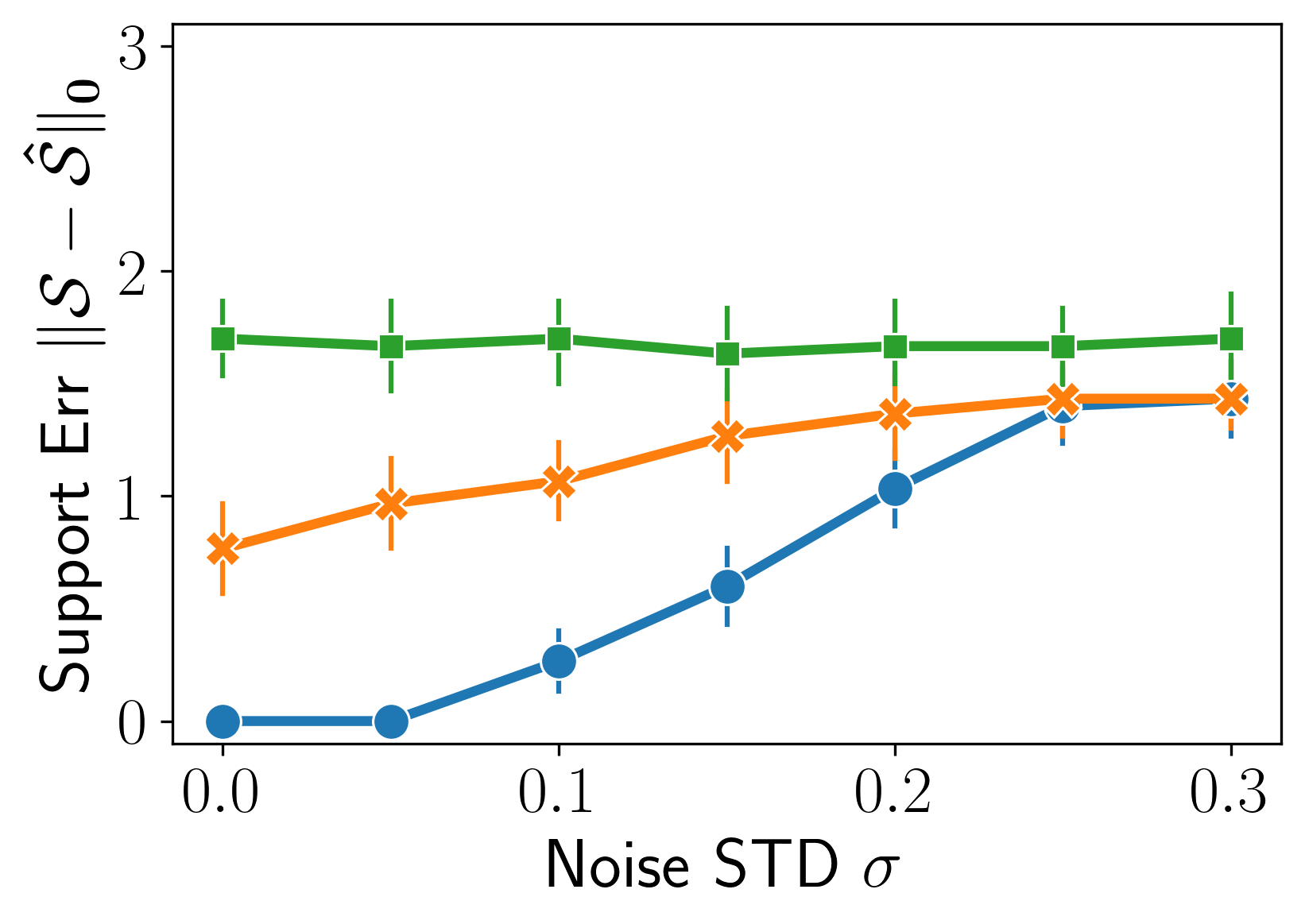}
         \caption{Cardinality $k=3$, rows $M=8$.}
         \label{fig:mot_rec_vs_noise}
     \end{subfigure}
     \hfill
     \begin{subfigure}[b]{0.329\textwidth}
         \centering
        \includegraphics[width=\textwidth]{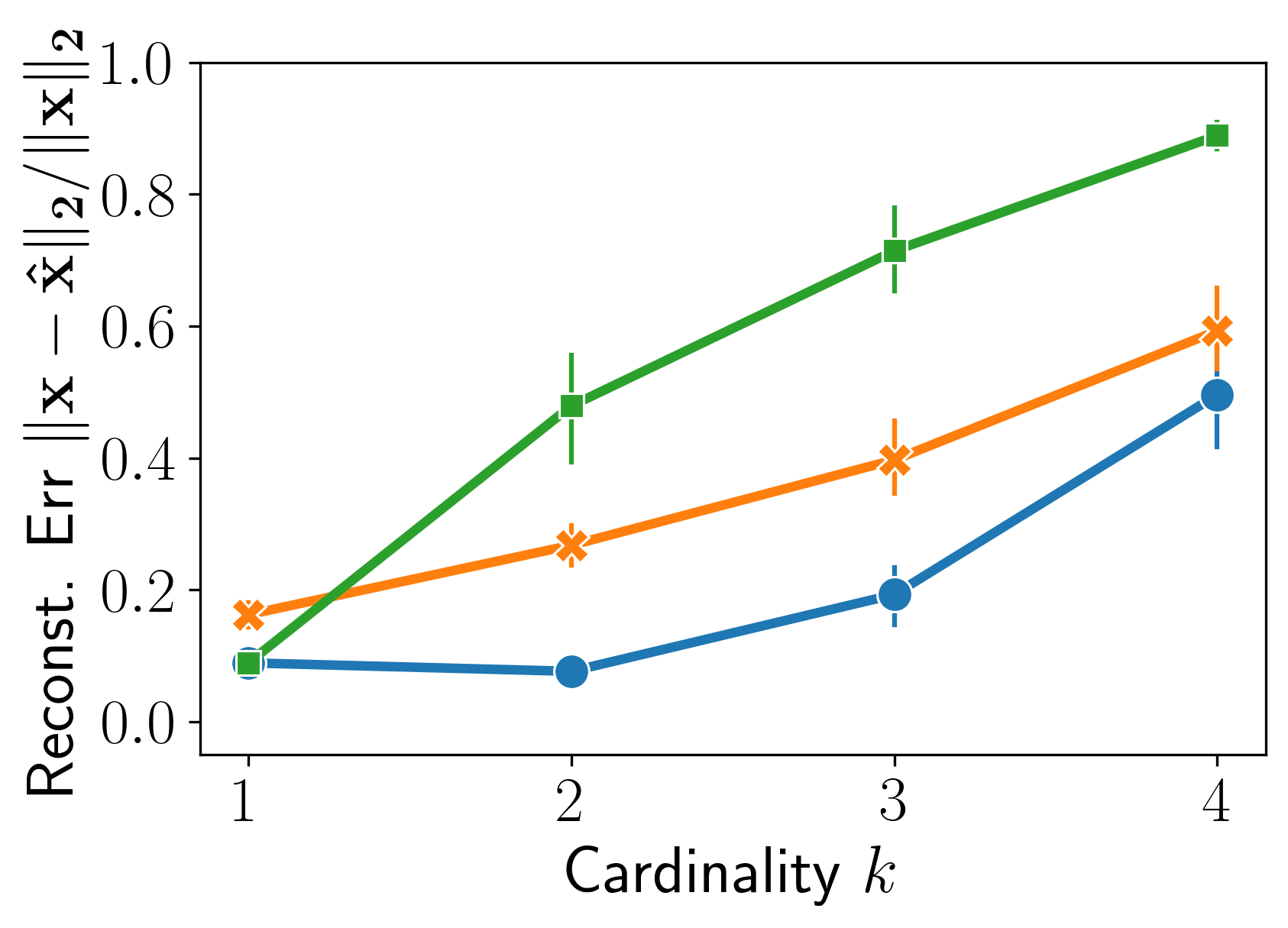}
         \includegraphics[width=\textwidth]{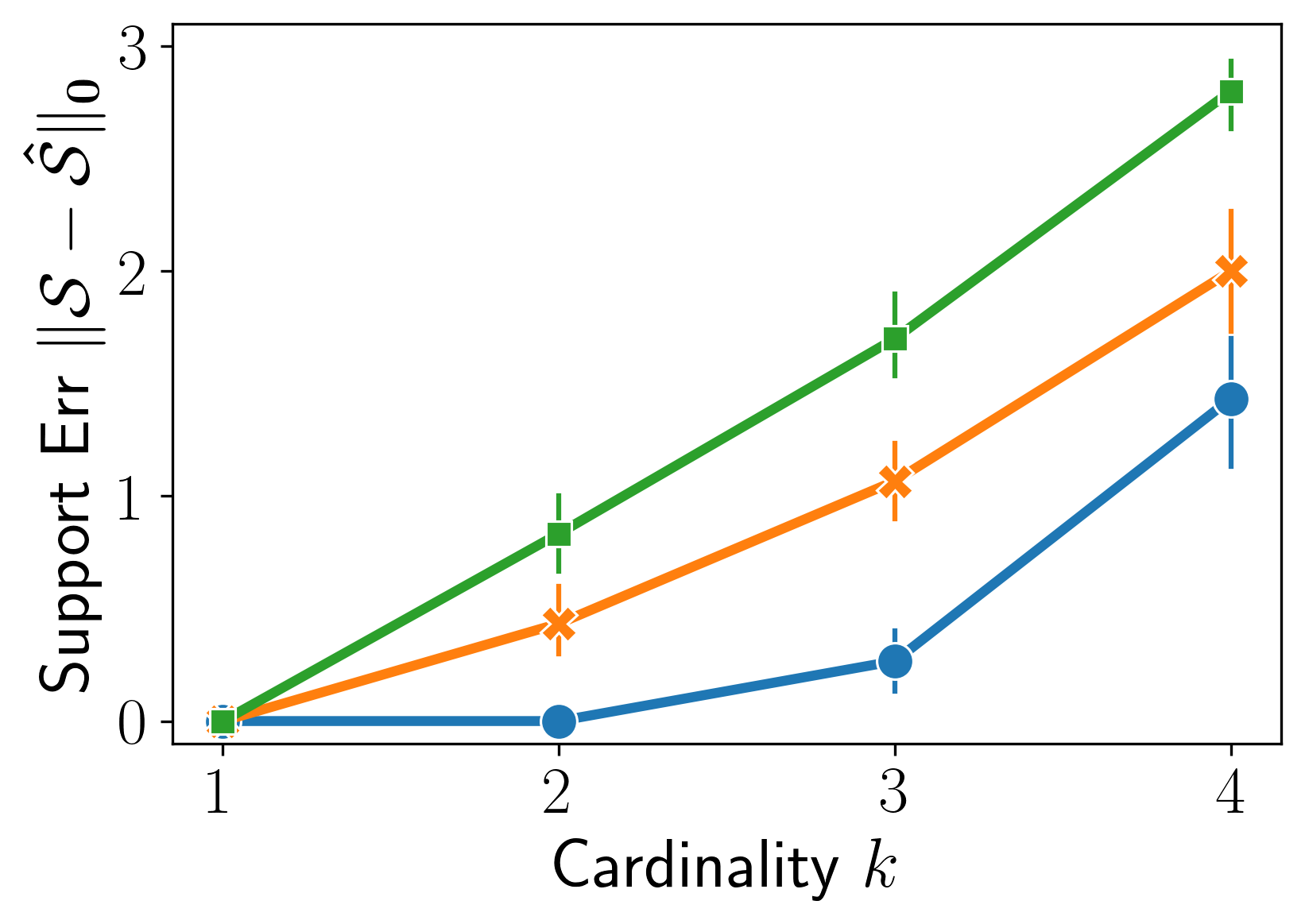}
         \caption{Rows $M=8$, noise $\sigma=0.1$.}
         \label{fig:mot_rec_vs_k}
     \end{subfigure}
     
        \caption{Comparison between the solutions obtained by Algorithm~\ref{alg:exact}, which exhaustively minimizes~\eqref{eq:k-sparse-problem}, and two classic approximation algorithms---lasso and OMP. (a)~Estimation error as a function the number of rows. (b)~Estimation error as a function the the noise level $\sigma$. (c)~Estimation error as a function of the number non-zeros $k$ in the true $\vx$. Each point in the graphs is an average over 30 independent realizations of $\mA,\vx$, and $\vv$; the length of $\vx$ is fixed and equals to  $N=16$.}
        \label{fig:motivation}
\end{figure}

Given such a matrix $\mA$, we generate a vector $\bar{\vb}=\mA\vx$ by drawing an $\vx$ with $k$ randomly chosen non-zero entries that are equal to~$1$, and then forming its noisy version $\vb = \bar{\vb} + \vv$, where the noise component $\vv$ is sampled from a Normal distribution with mean zero and standard deviation (STD) $\sigma$. Now, given $\mA$ and the noisy $\vb$, we attempt to reconstruct $\vx$ via lasso, OMP, and an exhaustive algorithm for minimizing~\eqref{eq:k-sparse-problem}, detailed in Algorithm~\ref{alg:exact}. Due to the high computational complexity of Algorithm~\ref{alg:exact}, we could run this experiment only for a matrix $\mA$ with a small the number of columns, which we set to $N=16$. The hyper-parameter of each method (the $L_1$ regularization strength in lasso, and the cardinality level in OMP and in Algorithm~\ref{alg:exact}) is tuned by an ``oracle'' that has access to the true $\vx$, optimizing each of the following two error metrics. The first metric evaluates the reconstruction error, defined as $\|\vx-\hat{\vx}\|_2/\|\vx\|_2$. The second measures the support recovery error, defined as $\|\mathcal{S} - \hat{\mathcal{S}}\|_0$, where $\mathcal{S} \in \mathbb{B}^N$ is a binary vector that contains the value $\mathcal{S}_i = 1$ in the $i$th index if $\vx_i \neq 0$, or the value $\mathcal{S}_i = 0$ otherwise. The vector $\hat{\mathcal{S}} \in \mathbb{B}^N$ is defined similarly, but with respect to the estimated vector $\hat{\vx}$. In plain words, the support recovery error is the number of non-zero elements in the unknown $\vx$ and estimated $\hat{\vx}$ that do not coincide in their locations.

Figure~\ref{fig:motivation} compares the performance of each method as a function of the sample size~$M$ (left), noise level~$\sigma$ (middle), and cardinality~$k$ (right).
Following that figure, we can see that the exhaustive method for minimizing \eqref{eq:k-sparse-problem} outperforms lasso and OMP by a large gain, especially (i)~in small sample size regimes, (ii)~for small noise levels, and (iii)~for higher cardinality values. This toy experiment, which explicitly shows how powerful is the original $L_0$ regularization compared to its surrogates, elucidates our strong desire to leverage quantum computing technologies to better approximate the NP-hard sparse coding problem, and further scale-up the dimension of $\mA$ to have a much larger number of columns.

\begin{algorithm}
\caption{An exhaustive solver for \eqref{eq:k-sparse-problem}.}\label{alg:exact}
\begin{algorithmic}
\Require A matrix $\mA$ of size $M \times N$, a noisy vector $\vb$ of length $M$, and a sparsity level $k$. 
\Ensure A $k$-sparse vector $\hat{\vx}$ minimizing \eqref{eq:k-sparse-problem}.

\State $S \gets $ a set containing all the possible binary support vectors with $k$ non-zero elements  \Comment{$N \choose k$ vectors}
\State $\hat{\vx} \gets \bm{0} $, where $ \bm{0}$ is an $N$-dimensional vector of zeros.
\State $R \gets \|\vb\|_2^2$. \Comment{Squared error of the zero-vector solution}
\For{all $\vs$ in $S$}
\State $\mA_{\vs} \gets$ a matrix of size $M \times k$ containing $k$ columns from $\mA$, specified by the support vector $\vs$.
\State $\hat{\vz}_{\vs} \gets (\mA_{\vs}^{\top}\mA_{\vs})^{-1}\mA_{\vs}^{\top} \vb $.   \Comment{Solution for the least squares problem: $ \min_{\vz} \| \mA_{\vs} \vz - \vb\|_2^2$}
\State ${r} \gets \| \mA_{\vs} \hat{\vz}_{\vs} - \vb\|_2^2$.
\If{${r} < R$} \Comment{The current solution is better that the best we found thus far}
    \State $R \gets r$.
    \State $\hat{\vx}_{\vs} \gets \hat{\vz}_s$, where the off-support elements that are not specified by $\vs$ are set to zero.
\EndIf
\EndFor
\end{algorithmic}
\end{algorithm}

\section{Quantum solvers for sparse coding problems}\label{sec:quantum-coding}
The problem of solving a system of linear equations with minimum errors can be reformulated as minimizing a quadratic function of binary variables, i.e., as a QUBO problem. The derivation presented in this section decomposes the objective function~\eqref{eq:lambda-problem} into the squared error term $\|\mA\vx-\vb\|_2^2$ and the sparsity penalty term $\|\vx\|_0$, forming the QUBO matrix for each component, denoted by $\mW^{L_2}$ and $\mW^{L_0}$, respectively. At a high level, the QUBO problem that we formalize takes the following form:
\begin{equation}
\label{eq:qubo-total}
    \min_{\vq \in \mathbb{B}^{D}} {\vq}^{\top} ({\mW}^{L_2} + \lambda {\mW}^{L_0}) \vq,
\end{equation}
where $\vq$ is a binary vector of spins. These spins express the unknown elements in the vector $\vx$, using fixed-point arithmetic as follows:
\begin{equation}
\label{eq:fp_rep}
    x_i = c_i^{\text{min}} + d_i \sum_{p=1}^P q_{i,p}2^{p-1}, \quad P \geq 1.
\end{equation}
Above, $c_i^{\text{min}}, d_i$ and $P$ are pre-specified constants: $c_i^{\text{min}}$ is the minimal value that can be expressed, $d_i$ is a scaling factor, and $P$ is the number of bits allocated; the variables $$q_{i,p} \in \mathbb{B}, \ \ 1 \leq i \leq N, \ 1 \leq p \leq P $$ are the spins, stored in the $NP$ dimensional vector
\begin{equation}
\label{eq:q}
    \vq = [q_{1,1}, \dots, q_{1,P}, q_{2,1}, \dots, q_{2,P}, \dots, q_{N,1}, \dots, q_{N,P}]^{\top}.
\end{equation}
The above is the most general representation for $\vx$. For example, consider the case where $c_i^{\text{min}}=-1$, $d_i=1$, and $P=2$. Here, $x_i$ can get one of the values in the set $\{-1, 0, 1, 2\}$, corresponding to the four possible combinations of $q_{i,p} \in \mathbb{B}, \ 1 \leq p \leq 2$. Importantly, throughout this paper, we require that the value $0$ is always contained in the possible values that $x_i$ can get; otherwise this feature would always be considered a non-zero. We shall note that the analysis of the most general fixed-point representation for $\vx$ with $P\geq3$ requires us to expand $\vq$ by adding one auxiliary spin per entry $x_i$ and revising \eqref{eq:qubo-total} accordingly; we discuss this in detail in Section~\ref{sec:qubo_l0}.




\subsection{QUBO matrix for the squared error term}
\label{sec:qubo_residual}

In this section, we focus only on the minimization of the squared error term $\|\mA\vx-\vb\|_2^2$ with respect to $\vx$, and show how to formulate it as a QUBO problem that minimizes the same objective, however, with respect to the spins $\vq$. To do so, we first expand the $L_2$ norm and rewrite it as follows:
\begin{equation}
\label{eq:res_x_space}
H_{L_2} := \|\mA\vx-\vb\|_2^2 = \sum_{i=1}^{N}\sum_{j=1}^{N}x_ix_j W_{i,j}^{\text{base,1}} + \sum_{i=1}^N x_i W_{i,i}^{\text{base,2}} + h^{\text{base}}, 
\end{equation}
where $\mW^{\text{base},1}$ is a matrix of size $N \times N$ with entries
\begin{align}
    & W_{i,j}^{\text{base},1} := \sum_{m=1}^M A_{m,i}A_{m,j}, \ \ 1 \leq i,j \leq N;
\end{align}
the diagonal matrix $\mW^{\text{base},2}$ is also of size $N \times N$ with diagonal elements
\begin{align}
    W_{i,i}^{\text{base},2} := -2 \sum_{m=1}^{M} A_{m,i}b_m, \ \ 1 \leq i \leq N;
\end{align}
and $h^{\text{base}}$ is a constant scalar, defined as
\begin{align}
h^{\text{base}} := \sum_{m=1}^M b_m^2.
\end{align}
Next, we plug into \eqref{eq:res_x_space} the explicit fixed-point expression for $x_i$ defined in~\eqref{eq:fp_rep}, expressing $H_{L_2}$ using the binary spins $q_{i,p} \in \mathbb{B}$ in lieu of $x_i$. After applying basic algebraic manipulations, we get the following expression:
\begin{equation}
\label{eq:res_fp_rep}
H_{L_2} = \sum_{i=1}^{N}\sum_{j=1}^{N}\sum_{s=1}^{P}\sum_{p=1}^{P} q_{i,s}q_{j,p} W_{s + P(i-1),p + P(j-1)}^{L_2,1} + \sum_{i = 1}^N \sum_{p = 1}^P q_{i,p} W_{p + P(i-1),p + P(i-1)}^{L_2,2} + h^{L_2}.
\end{equation}
Above, $\mW^{L_2,1}$ is a matrix of size $NP \times NP$ whose entries are given by
\begin{align}
    W_{s + P(i-1),p + P(j-1)}^{L_2,1} := 2^{s+p-2}W_{i,j}^{\text{base},1}d_id_j, \ \ 1 \leq i,j \leq N, \ \ 1 \leq s,p \leq P.
\end{align}
In addition, $\mW^{L_2,2}$ is an $NP \times NP$ diagonal matrix whose diagonal elements are expressed as
\begin{align}
    W_{p + P(i-1),p + P(i-1)}^{L_2,2} := 2^{p-1} d_i \left( W_{i,i}^{\text{base},2} + 2 \sum_{j=1}^{N} c_j^{\text{min}} W_{i,j}^{\text{base},1} \right), \ 1 \leq i \leq N, \ \ \ \ 1 \leq p \leq P,
\end{align}
and the constant $h^{L_2}$ is formulated as
\begin{align}
    h^{L_2} := \sum_{i=1}^N\sum_{j=1}^N W_{i,j}^{\text{base},1}c_i^{\text{min}}c_j^{\text{min}} + \sum_{i=1}^N W_{i,i}^{\text{base},2} c_i^{\text{min}} + h^{\text{base}}.
\end{align}
Now, we make two observations. First, since $q_{i,p}$ is binary we have $q_{i,p} q_{i,p} = q_{i,p}^2 = q_{i,p} $, and therefore we can add the diagonal elements of $\mW^{L_2,1}_{ip,ip}$ to those of $\mW^{L_2,2}_{ip,ip}$. Second, the minimization of $H_{L_2}$ with respect to $\vq$ is not affected by $h^{L_2}$ since the latter is a constant. These two observations complete our derivation of the QUBO matrix for the squared error term in~\eqref{eq:lambda-problem}, concluding that the matrix $\mW^{L_2}$ in \eqref{eq:qubo-total} is given by
\begin{equation}
\label{eq:W_L2}
    \mW^{L_2} = \mW^{L_2,1} + \mW^{L_2,2}.
\end{equation}

\subsection{QUBO matrix for the cardinality term}
\label{sec:qubo_l0}
We turn to develop an explicit expression for the QUBO matrix that corresponds to the sparsity penalty. The idea is to compute the cardinality of the estimated $\vx$ under the fixed-point representation, implemented via the binary vector $\vq$. Since we allow $x_i$ to have negative values, we use the following strategy to facilitate the representation of the zero elements in $\vx$. Denote the combination of the binary elements that leads to $x_i=0$ by $c_{i,p}^0$, satisfying
\begin{equation}
    x_i = 0 = c_i^{\text{min}} + d_i \sum_{p=1}^P c_{i,p}^02^{p-1}.
\end{equation}
Recall our example where we set $c_i^{\text{min}}=-1$, $d_i=1$, and $P=2$, in which $x_i\in\{-1,0,1,2\}$. In this case, the value $x_i=0$ is obtained for $c^0_{i,1}=1$ and $c^0_{i,2}=0$. Importantly, the  binary elements $c^0_{i,p}$ are known constants, as these are derived from the pre-determined constants $c_i^{\text{min}}$, $d_i$, and $P$.
Given the constants $c^0_{i,p}$, we define
\begin{equation}
\label{eq:transformed_spins}
    y_{i,p} = \begin{cases}
    1 - q_{i,p}, &c_{i,p}^0 = 0, \\
    q_{i,p}, &c_{i,p}^0 = 1,
    \end{cases} 
\end{equation}
where the binary variables $y_{i,p}\in\mathbb{B}$ can be thought of as ``transformed spins''. Importantly, under the above formulation, we have
\begin{align}
    y_{i,p}=1  \ \ \forall 1\leq p \leq P \ \ \xleftrightarrow \ \  
    c^0_{i,p}=q_{i,p}  \ \ \forall 1\leq p \leq P \ \ \xleftrightarrow \ \ 
    x_i = 0,
\end{align}
where $\leftrightarrow$ stands for ``if and only if''.
In words, the above implies that $x_i$ is equal to zero if and only if all the transformed spins $y_{i,p}$, $1 \leq p \leq P$ are `active' and equal to $1$.  Compactly, this can be expressed as
\begin{equation}
\label{eq:zi}
    z_i = y_{i,1}y_{i,2} \dots y_{i,P},
\end{equation}
where $z_i = 1$ if and only if $x_i=0$. Now, we can compute how many non-zeros we have in $\vx$ by
\begin{equation}
    \label{eq:L0_H}
    H_{L_0} := \|\vx\|_0 = \sum_{i=1}^N(1-z_i).
\end{equation}
Therefore, the formulation of the transformed spins $y_{i,p}$ allows us to express the sparsity penalty in QUBO terms, for the most general fixed-point representation. Unfortunately, in this general case, the variables $z_i$ are not quadratic in the spins $q_{i,p}$, and therefore \eqref{eq:L0_H} can not be used in its present form to construct the matrix $\mW^{L_0}$ in \eqref{eq:qubo-total}. Yet, we will show how to overcome this challenge using auxiliary ancilla spins. 
Before doing so, however, we pause to discuss two important special cases. The first is a case where $\vx$ is binary with $P=1$ and $d_i=1$ in~\eqref{eq:fp_rep}, and the second extends the former to the 2-bit case, with $P=2$. In contrast to the $P\geq3$ case, both the binary and 2-bit representation do not require the use of ancilla spins.

\subsubsection*{Binary representation: $\mathbold{P}$=1} 
In this simple setting, $x_i = q_i$ by construction, as $q_i$ is the binary spin that corresponds to the $i$th entry in the binary $\vx$. Therefore, the number of non-zeros in the estimated $\vx$ is nothing but the sum over all the original spins, which can be written as follows:
\begin{equation}
    H_{L0}^{\text{binary}} = \|\vx\|_0= \sum_{i=1}^N q_i = \sum_{i=1}^N q_iq_i = \vq^{\top}\mW^{L_0}\vq.
\end{equation}
Above, the third equality holds since $q_i\in\mathbb{B}$, and the forth equality holds for the identity matrix $\mW^{L_0} = \mI$ of size $N \times N$.

\subsubsection*{2-bit representation: $\mathbold{P}$=2} 

In contrast to the binary case, in this setting the cardinality of $\vx$ is not equal anymore to the sum the original spins $q_{i,p}$, $1\leq i \leq N$, $1 \leq p \leq 2$. Yet, following~\eqref{eq:L0_H}, we can express the sparsity penalty using the two transformed spins $y_{i,1}$ and $y_{i,2}$, where this transformation is pre-defined via the constants $c^0_{i,p}\in\mathbb{B}$, $1 \leq p \leq 2$. 
Below, we show that for the special case of $P=2$, the expression in~\eqref{eq:L0_H} is a quadratic function of the original spins $q_{i,p}$, which is crucial to yield the QUBO matrix $\mW^{L_0}$. The idea is to map the transformed spins $y_{i,p}$ back to the original space of $q_{i,p}$, for all $i,p$, and then rewrite the resulting expression in a form of $\vq^\top \mW^{L_0} \vq$. Concretely, consider our running example, in which $c_i^{\text{min}}=-1$, $d_i=1$, and $P=2$. Here, $x_i$ can get negative values, however we know that $x_i=0$ for the constants $c^0_{i,1}=1$ and $c^0_{i,2}=0$. Now, imagine we start with a $2N\times 2N$ matrix $\mW^{L_0}$ of zeros, and we wish to fill-in the entries in $\mW^{L_0}$ that correspond to $x_i$. According to~\eqref{eq:transformed_spins}, since $c_{i,1}^0=1, \ c_{i,2}^0=0$, we have 
\begin{equation}
    y_{i,1}=q_{i,1} \ \text{and} \  y_{i,2}=1-q_{i,2},
\end{equation}
leading to $1-z_i$ that is quadratic in the spins $q_{i,p}$, since
\begin{equation}
\label{eq:2bits_example}
    1-z_i = 1-y_{i,1}y_{i,2} = 1-q_{i,1}(1-q_{i,2}) = 1 - q_{i,1}q_{i,1} + q_{i,1}q_{i,2}.
\end{equation}
The above implies that we should set the corresponding entries in the QUBO matrix as follows: 
\begin{equation}
\label{eq:W_L0_binary}
W^{L_0}_{1+2(i-1),1+2(i-1)} = -1; \  W^{L_0}_{1+2(i-1),2+2(i-1)} = \frac{1}{2}; \ \text{and} \  W^{L_0}_{2+2(i-1),1+2(i-1)} = \frac{1}{2}.
\end{equation}
Note that we ignore the leading constant `1' in the right-hand-side of~\eqref{eq:2bits_example} since constants do not affect the solution of the underlying optimization problem. The remaining elements in the matrix can be determined by repeating a similar set of steps for all $1 \leq i \leq N$, however with possibly  different values of $c^0_{i,1}, c^0_{i,2}$. Notice that there are four cases for $c^0_{i,p}\in\mathbb{B}$, $1 \leq p \leq 2$ that should be considered, while above we studied only one of these for which $c^0_{i,1}=1$ and $c^0_{i,2}=0$. For completeness, we present below the three remaining cases. For $c_{i,1}^0=0, \ c_{i,2}^0=0$, we follow \eqref{eq:transformed_spins} and get $y_{i,1}=1-q_{i,1} \ \text{and} \  y_{i,2}=1-q_{i,2}$, leading to $1-z_i = q_{i,1} + q_{i,2} - q_{i,1}q_{i,2}$. Analogously, when $c_{i,1}^0=0, \ c_{i,2}^0=1$ we have $y_{i,1}=1-q_{i,1} \ \text{and} \  y_{i,2}=q_{i,2}$, leading to $1-z_i = 1-q_{i,2} + q_{i,1}q_{i,2}$. The last case deals with
$q_{i,1}^0=1, \ q_{i,2}^0=1$; here, $y_{i,1}=q_{i,1} \ \text{and} \  y_{i,2}=q_{i,2}$, and thus $1-z_i = 1- q_{i,1}q_{i,2}$. It is immediate to explicitly express $\mW^{L_0}$ for each case as in~\eqref{eq:W_L0_binary}; we omit this in the interest of space.

\subsubsection*{The general case: $\mathbold{P\geq}$ 3} 

This is the most difficult case to address, as a naive extension of \eqref{eq:zi} to the $P$-bit case would result in $H_{L_0}$ in \eqref{eq:L0_H} that has high-order interactions between the original spins, breaking the bilinear structure of the QUBO problem. As a way out, we introduce ancilla spins, which allow us to express $H_{L_0}$ for the most general representation of $\vx$ in a quadratic form that perfectly fits the QUBO structure. We take inspiration from \cite{freedman2005energy,ishikawa2010transformation} and offer a solution that is extremely efficient with respect to the number of ancilla spins: we add only one auxiliary spin per feature, which we view as the minimal number of spins that one can hope for, in such a general case. 

Turning to the details, denote the ancilla spin for $x_i$ by $s_{i}\in\mathbb{B}$ and define the function
\begin{equation}
\label{eq:F}
    F(y_{i,1}, y_{i,2}, \dots, y_{i,P}, s_i) := s_i \cdot \left(y_{i,1} + y_{i,2} + \dots + y_{i,P} - (P - 1)\right),
\end{equation}
which gets as input all the transformed spins $y_{i,p}$ as well as the ancilla spin $s_i$ of $x_i$. Now, we invoke a beautiful result presented in \cite{freedman2005energy} and \cite[Section 4.2]{ishikawa2010transformation}, stating that
\begin{equation}
    -z_i = -y_{i,1}y_{i,2} \dots y_{i,P} = \min_{s_i} -F(y_{i,1}, y_{i,2}, \dots, y_{i,P}, s_i).
\end{equation}
In plain words, the minimal value of the function $-F(y_{i,1}, y_{i,2}, \dots, y_{i,P}, s_i)$, optimized with respect to the ancilla spin $s_i$, is in fact equal to $-z_i$. Since the function $F$ in \eqref{eq:F} is a sum of $s_iy_{i,p}$, $1 \leq p \leq P$, we can harness it to form a cost function that is equivalent to $H_{L_0}$ in \eqref{eq:L0_H}, but involves only bilinear spin terms. Specifically, we define our cardinality Hamiltonian as
\begin{align}
\label{eq:general_HL0_F}
    H_{L_0} = \|\vx\|_0 = \sum_{i=1}^N(1-z_i) &= \sum_{i=1}^N(1-F(y_{i,1}, y_{i,2}, \dots, y_{i,P})) \\
    &= \sum_{i=1}^N\left(1-s_iy_{i,1}-s_iy_{i,2}-\dots-s_iy_{i,P} + s_i\left(P-1\right)\right).
\end{align}
Armed with the above cost function, we are able to form the structure and content of the matrix $\mW^{L_0}$. One technical challenge is that $y_{i,p}$ are the transformed versions of the original spins $q_{i,p}$, and so we must substitute the spins $q_{i,p}$ directly into \eqref{eq:general_HL0_F}. This step is similar in spirit to the 2-bit case that we have already discussed in depth earlier. By doing so, we obtain an expression that is quadratic with respect to the $NP$ original spins $q_{i,p}$, as well as with the additional $N$ ancilla spins $s_i$. Concretely, we define the vector $\tilde{\vq}=[\vq \ ; \ \vs]^{\top}$ that contains a total of $N(P + 1)$ spins; the first $NP$ elements are the original spins $\vq$, and the rest are the $N$ ancilla spins
$$ \vs = [s_1, s_2, \dots, s_{N}].$$
Consequently, the matrix $\mW^{L_0}$ is of size $N(P + 1) \times N(P + 1)$, where we use the convention that the first $NP$ rows (resp. columns) correspond to the spins $q_{i,p}$ and the rest $N$ rows (resp. columns) correspond to the ancilla spins $s_{i}$. Here, the QUBO problem defined with the vector of spins $\tilde{\vq}$ is given by
$$  \min_{\tilde{\vq} \in \mathbb{B}^d} \tilde{\vq}^{\top} (\widetilde{\mW}^{L_2} + \lambda \mW^{L_0}) \tilde{\vq},$$
where $\widetilde{\mW}^{L_2}$ is a matrix of size $N(P+1) \times N(P+1)$, obtained by padding the matrix $\mW^{L_2}$ in~\eqref{eq:W_L2} with zeros in all the entries that correspond to the ancilla spins.

To better understand how to construct $\mW^{L_0}$, it may be best to consider an example with $P=4$. Observe that each of the four transformed spins $y_{i,1},y_{i,2},y_{i,3},y_{i,4}$ can get one of two possible expressions, determined by the constants $c_{i,1}^0$ and $c_{i,2}^0$, $c_{i,3}^0$, $c_{i,4}^0$, respectively. This creates 16 different explicit forms for the term $F(y_{i,1},y_{i,2}, y_{i,3}, y_{i,4}, s_{i})$ as a function of $c_{i,1}^0,c_{i,2}^0,c_{i,3}^0,c_{i,4}^0$.
For instance, following \eqref{eq:transformed_spins}, the choice of $c_{i,1}^0=1,  c_{i,2}^0=0, c_{i,3}^0=0, c_{i,4}^0=1$ implies that $y_{i,1}=q_{i,1}  \ y_{i,2}=1-q_{i,2}, \ y_{i,3}=1-q_{i,3}, \ y_{i,4}=q_{i,4}$. Plugging the latter into the explicit expression of $F$ results in
\begin{align}
    1-z_i = 1-F(y_{i,1},y_{i,2},y_{i,3},y_{i,4}, s_{i}) &= 1-F(q_{i,1},1-q_{i,2}, 1-q_{i,3}, q_{i,4}, s_{i,1} ) \\
    &= 1-s_i (q_{i,1} + (1-q_{i,2}) + (1-q_{i,3}) + q_{i,4} - (4 - 1)) \\
    &=1 - s_i (q_{i,1} - q_{i,2} -q_{i,3} + q_{i,4} - 1).
    \label{eq:F_1_0_0_1}
\end{align}
Observe that the minimal value of \eqref{eq:F_1_0_0_1} is 0, obtained for the combination $q_{i,1}=c_{i,1}^0=1, q_{i,2}=c_{i,2}^0=0, q_{i,3}=c_{i,3}^0=0$, $q_{i,4}=c_{i,4}^0=1$, and $s_i=1$. Now, since the leading constant `1' in \eqref{eq:F_1_0_0_1} does not affect the combination of spins that minimizes this expression, we can set the entries in $\mW^{L_0}$ that correspond to \eqref{eq:F_1_0_0_1} as follows:
$$W^{L_0}_{1+4(i-1),N+i} = -1, \ W^{L_0}_{2+4(i-1),N+i} = 1, \ W^{L_0}_{3+4(i-1),N+i} = 1, \ W^{L_0}_{4+4(i-1),N+i} = -1, \ \text{and} \ W^{L_0}_{N+i,N+i} = +1.$$
A similar analysis can be done for the remaining 15 cases, which we do not provide here in the interest of space. 

In closing, this section introduced a novel decomposition of the sparse coding problem~\eqref{eq:lambda-problem} as a QUBO model. To formulate an efficient solution with respect to the number of spins, our analysis is separated into three cases: binary with $P=1$, 2-bit with $P=2$, and a general one with $P\geq3$, where the total number of spins varies with $P$. For $P=1$ we have $N$ spins, for $P=2$ we have $2N$ spins, and for the general $P\geq3$ case we have $N(P+1)$ spins in total.

\section{Experiments}\label{sec:res}

    \begin{figure}
     \centering
     \begin{subfigure}[b]{0.329\textwidth}
         \centering
         \includegraphics[width=\textwidth]{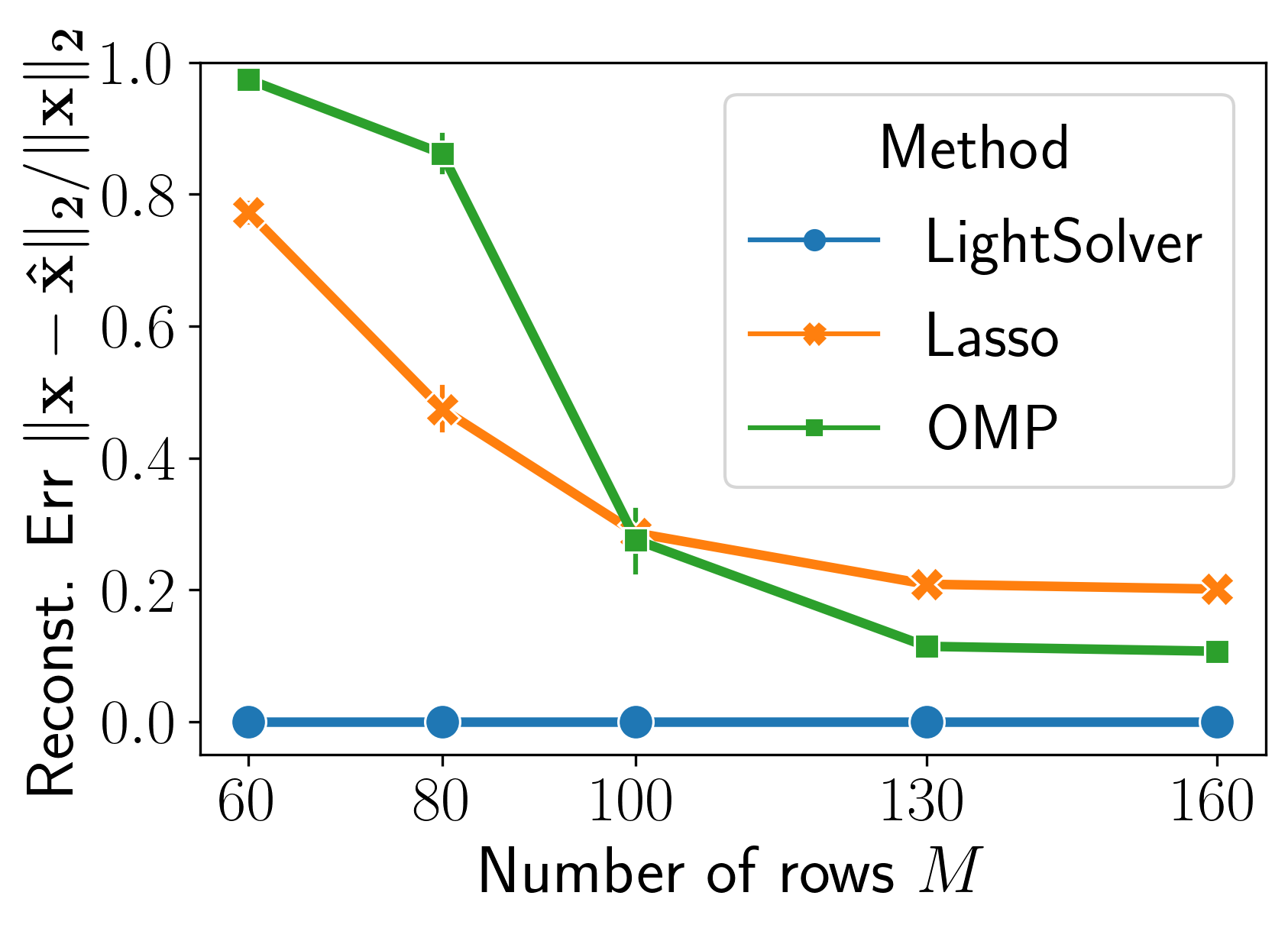}
         \includegraphics[width=\textwidth]{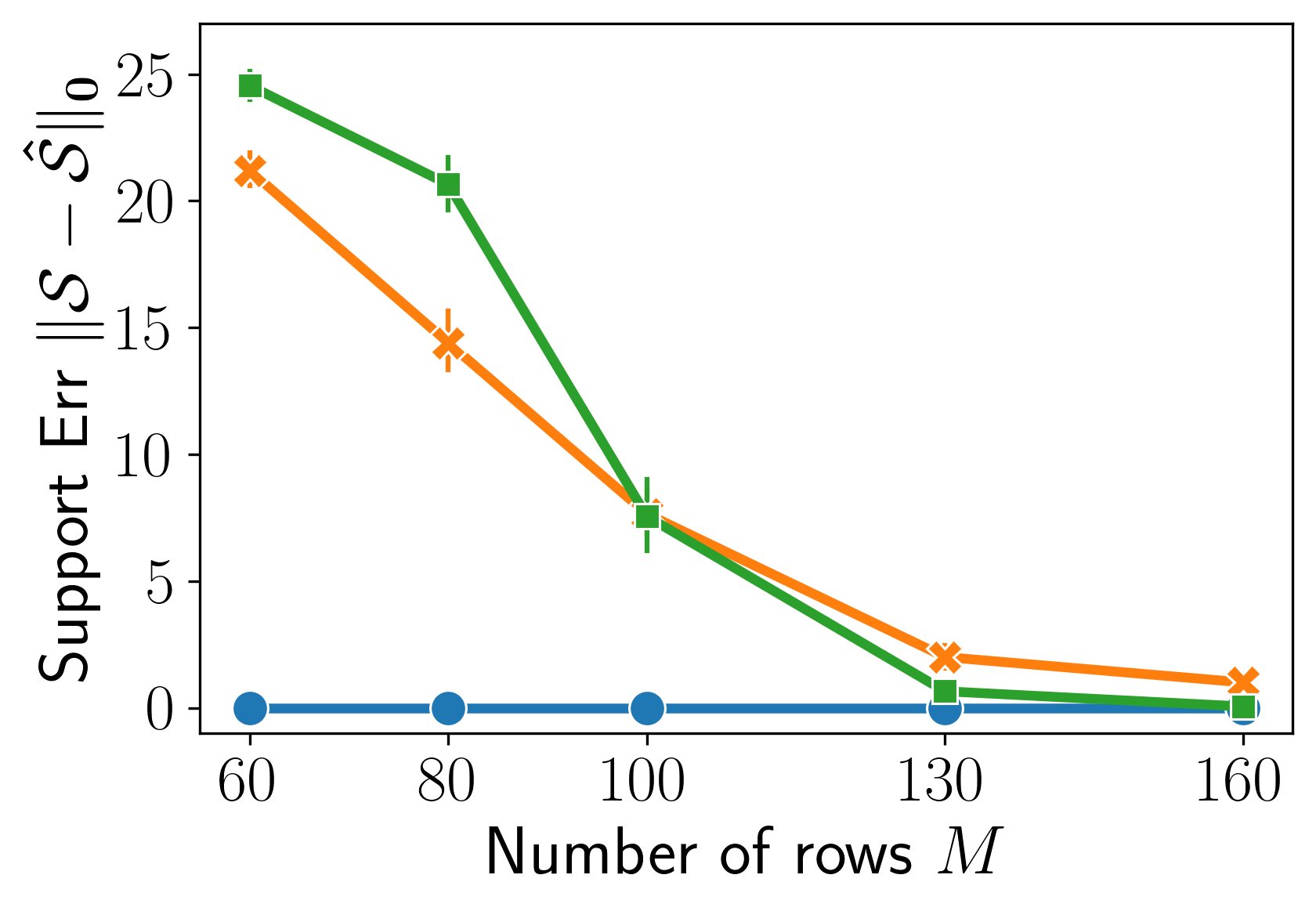}
         \caption{Cardinality $k=30$, noise $\sigma=0.1$.}
         \label{fig:bin_rec_vs_m}
     \end{subfigure}
     \hfill
     \begin{subfigure}[b]{0.329\textwidth}
         \centering
         \includegraphics[width=\textwidth]{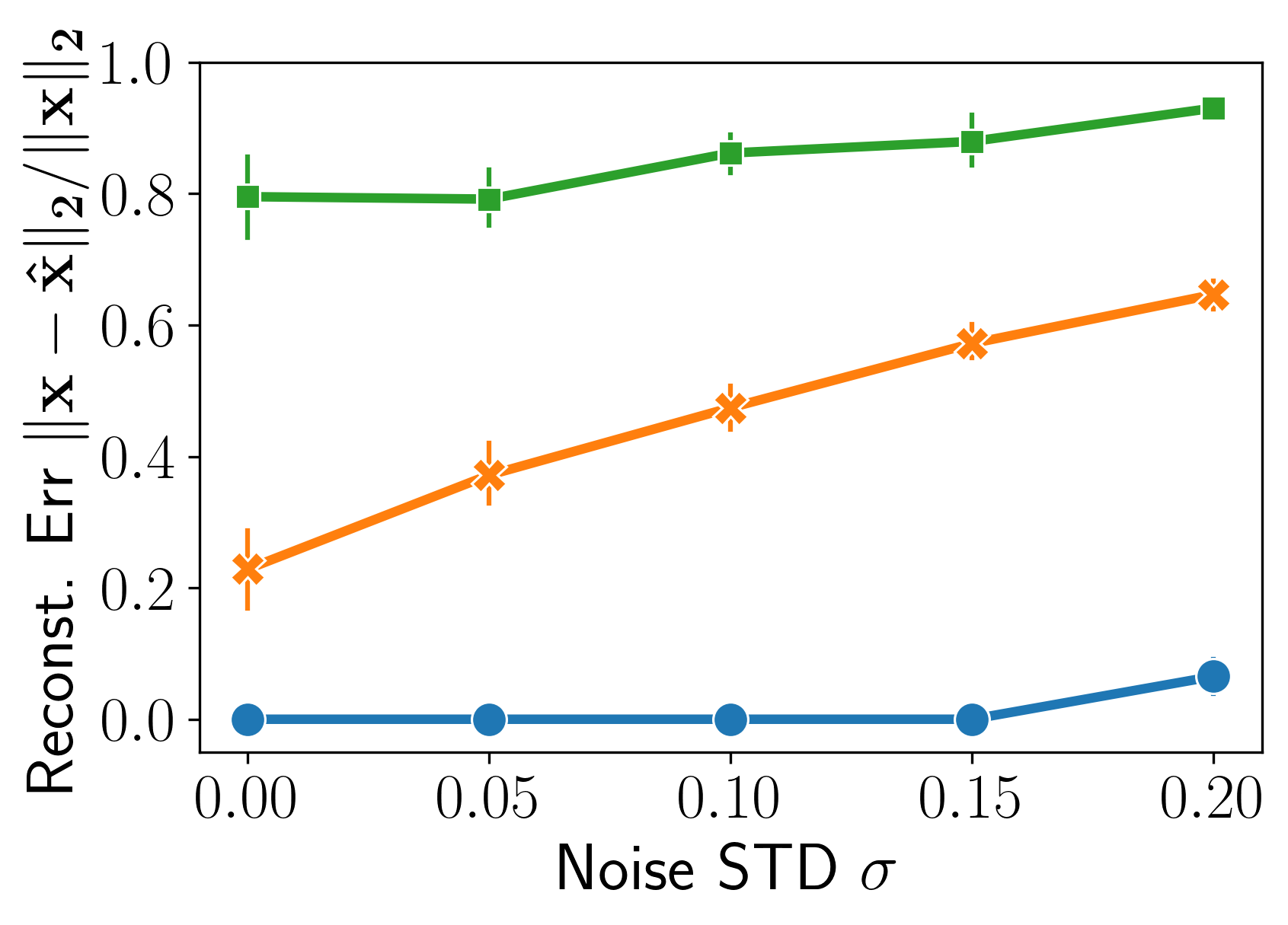}
         \includegraphics[width=\textwidth]{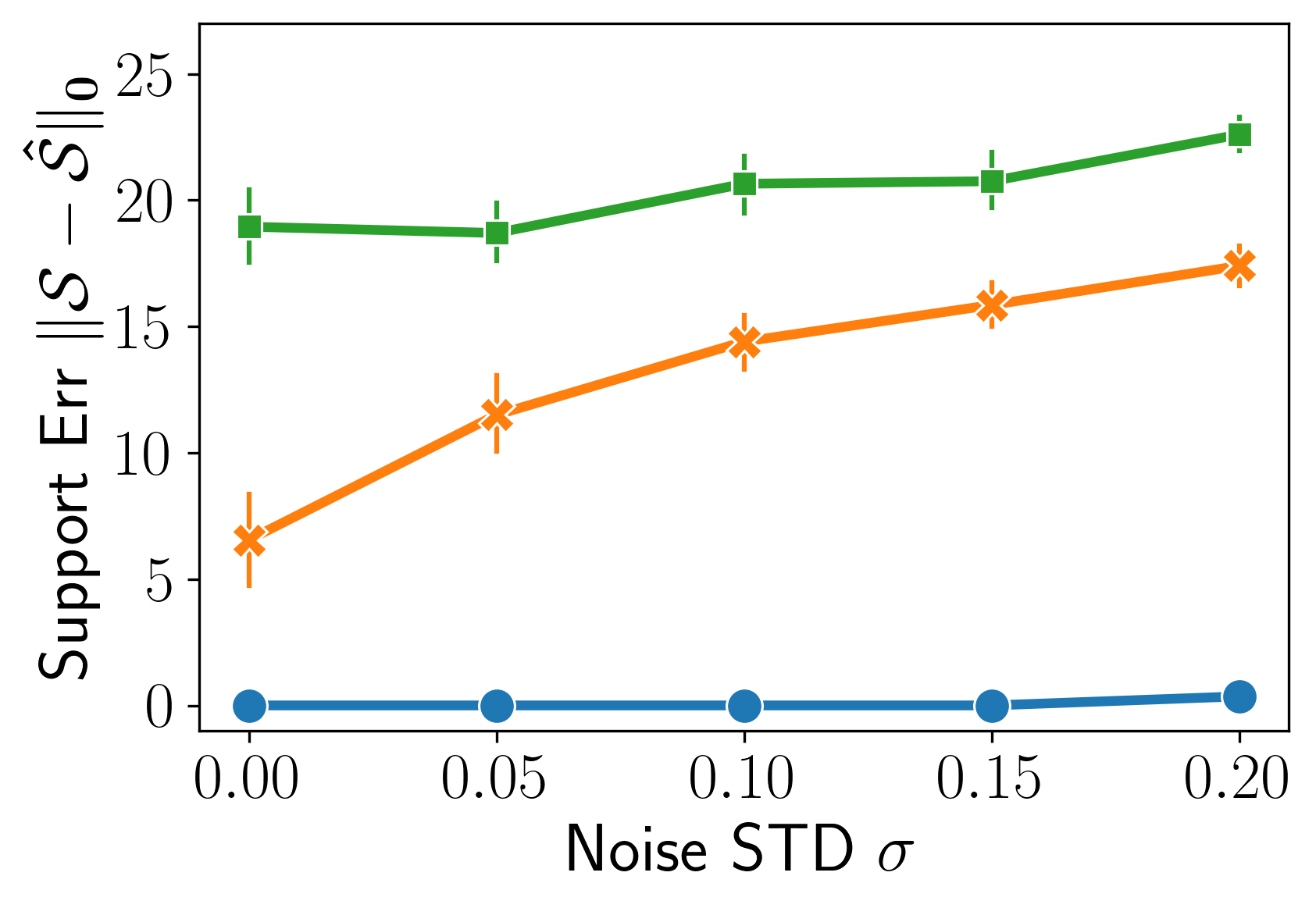}
         \caption{Cardinality $k=30$, rows $M=80$.}
         \label{fig:bin_rec_vs_noise}
     \end{subfigure}
     \hfill
     \begin{subfigure}[b]{0.329\textwidth}
         \centering
        \includegraphics[width=\textwidth]{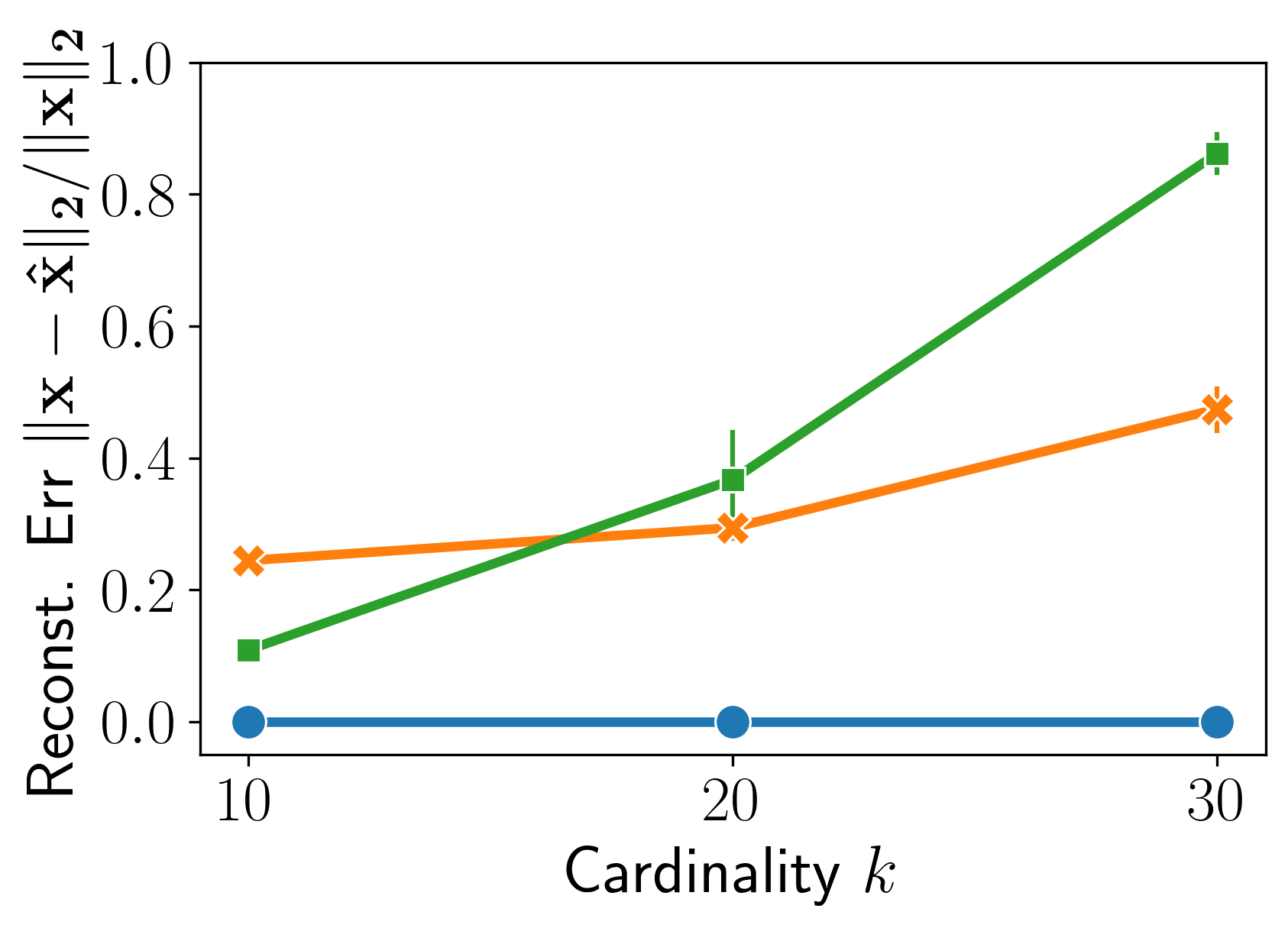}
         \includegraphics[width=\textwidth]{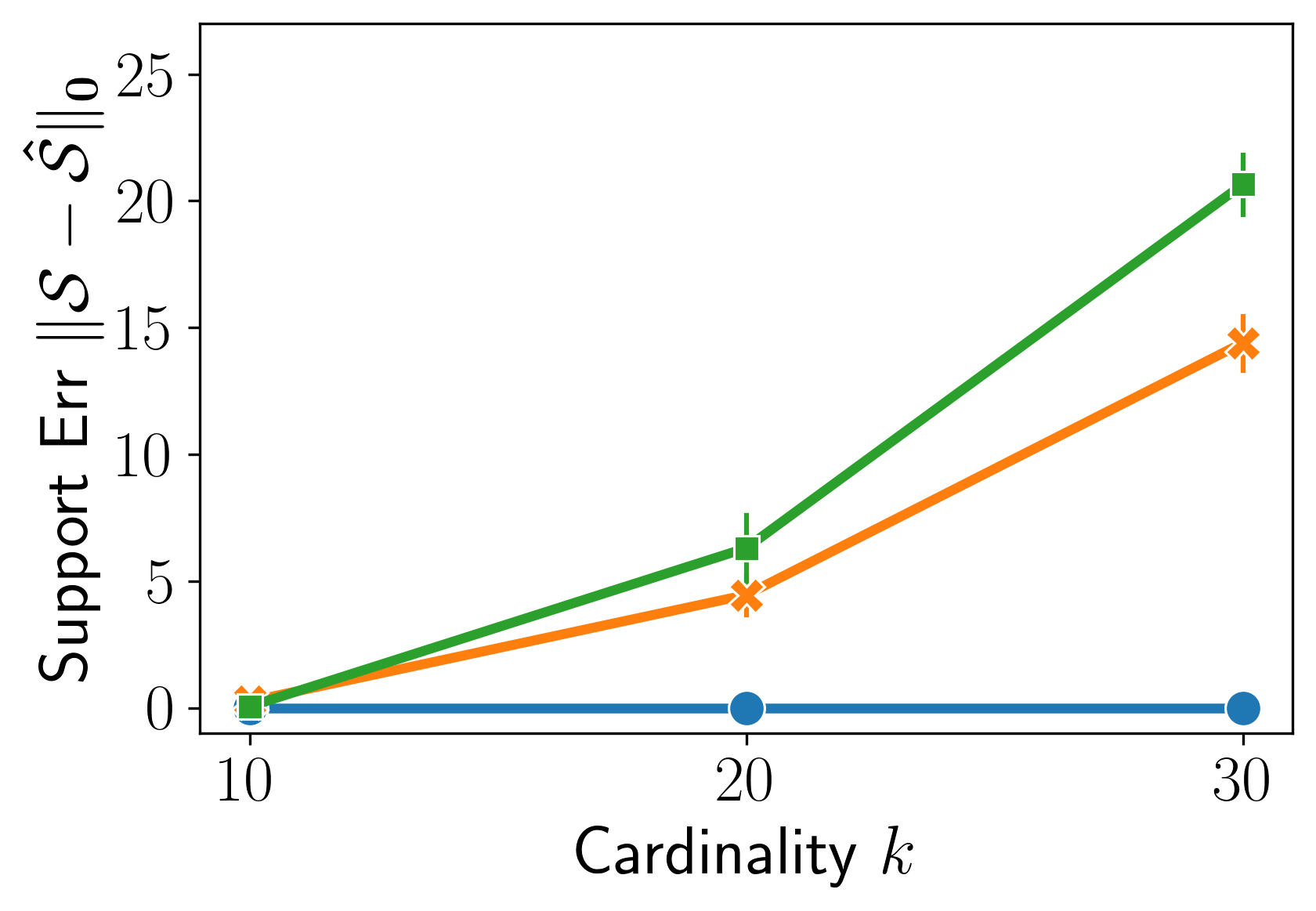}
         \caption{Rows $M=80$, noise $\sigma=0.1$.}
         \label{fig:bin_rec_vs_k}
     \end{subfigure}
     
        \caption{Results obtained by LightSolver's digital simulator, minimizing the proposed QUBO problem for a binary $\vx$ of length $N=160$. Each point in the graphs is an average over 20 independent realizations of $\mA,\vx$, and $\vv$. Other details are as in Figure~\ref{fig:motivation}.}
        \label{fig:binary}
\end{figure}

\paragraph{Binary sparse vectors.} We return to the experiment from Section~\ref{sec:motivation}, but now increase significantly the dimensions of the problem. We scale-up the dimensions of $\mA$ and $\vx$ by a factor of 10, focusing on a sparse \emph{binary} vector $\vx$ ($P=1$) of dimension $N=160$. The matrix $\mA$ is designed to have low mutual coherence according to~\eqref{eq:coherence}, generated as described in~Appendix~\ref{app:genA}. Since the vector $\vx$ is binary, we use the QUBO formulation with $P=1$, where the total number of spins is equal to~$N=160$. Armed with $\mW^{L_2}$ and $\mW^{L_0}$, we minimize~\eqref{eq:qubo-total} using a quantum-inspired annealer implemented by LightSolver's digital simulator. We then compare the quality of our quantum-inspired estimation $\hat{\vx}$ to lasso and OMP, serving as baseline methods.
Naturally, we cannot include Algorithm~\ref{alg:exact} as a baseline, since it is infeasible to run such an exhaustive optimization method with the dimensions of the problem that we study here. Similarly to Section~\ref{sec:motivation}, the hyper-parameter $\lambda$ of our method, as well as the hyper-parameters of lasso and OMP, are chosen by an ``orcale'' that has access to $\vx$, seeking the parameter that achieves the smallest possible error metric under study, for each method.
Figure~\ref{fig:binary} presents the reconstruction and support recovery errors as a function of $M$ (Figure~\ref{fig:bin_rec_vs_m}), $\sigma$ (Figure~\ref{fig:bin_rec_vs_noise}), and $k$ (Figure~\ref{fig:bin_rec_vs_k}). Observe how the quantum-inspired estimations are highly accurate, portraying nearly perfect recovery in all cases. In particular, the quantum-based estimations are superior than those obtained by the widely-used lasso and OMP, all across the board. This dramatic improvement can be explained as follows. First, the annealer accurately minimizes the NP-hard QUBO problem of interest. 
Second, lasso and OMP do not leverage the prior knowledge that $\vx$ is binary, in contrast to the annealer that utilizes this knowledge, encapsulated \textit{apriori} in the QUBO matrices.

\begin{figure}
     \centering
     \begin{subfigure}[b]{0.329\textwidth}
         \centering
         \includegraphics[width=\textwidth]{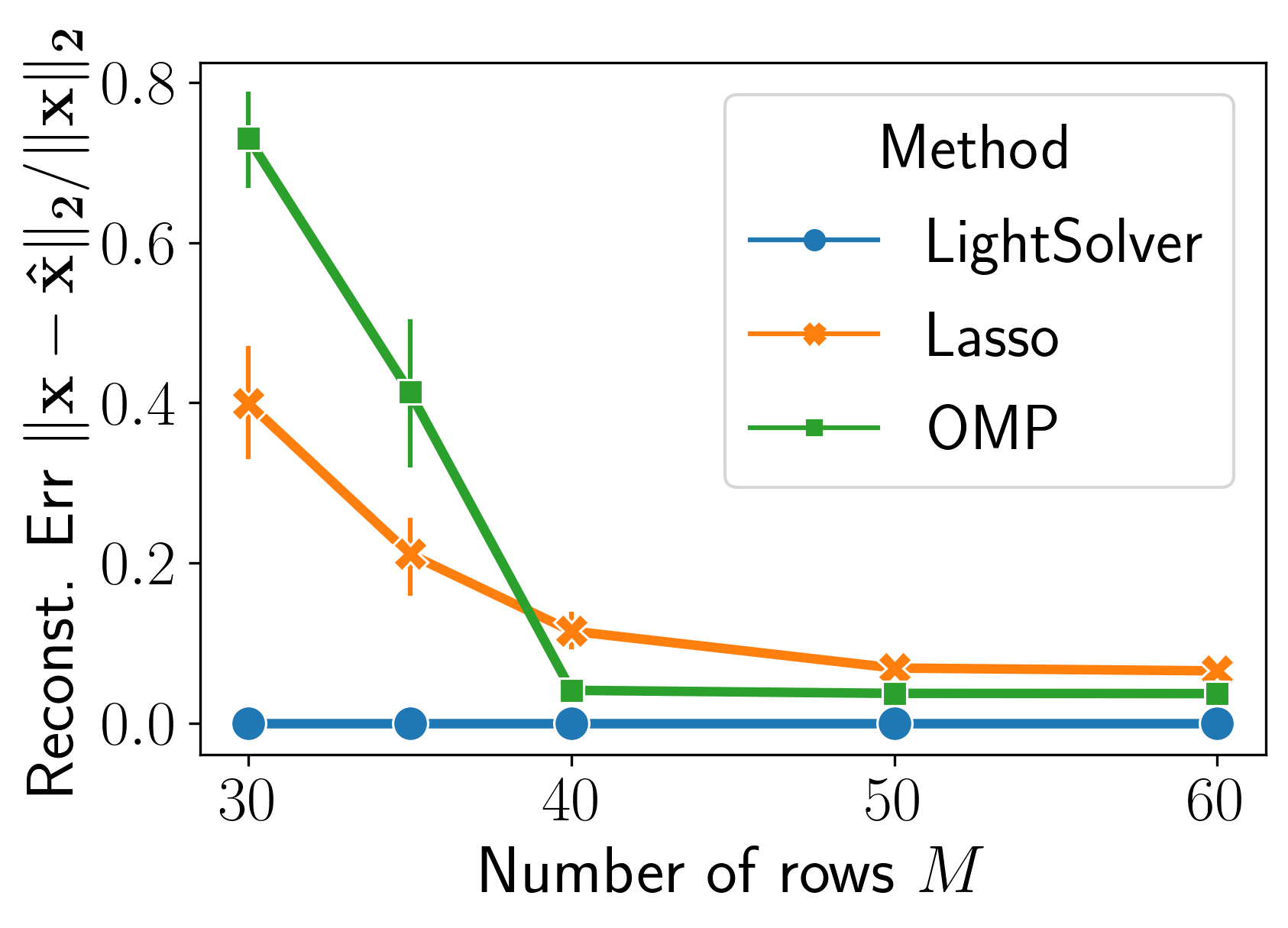}
          \label{fig:fp_rec_vs_m}
     \end{subfigure}
     \hspace{0.1cm}
     \begin{subfigure}[b]{0.329\textwidth}
         \centering
         \includegraphics[width=\textwidth]{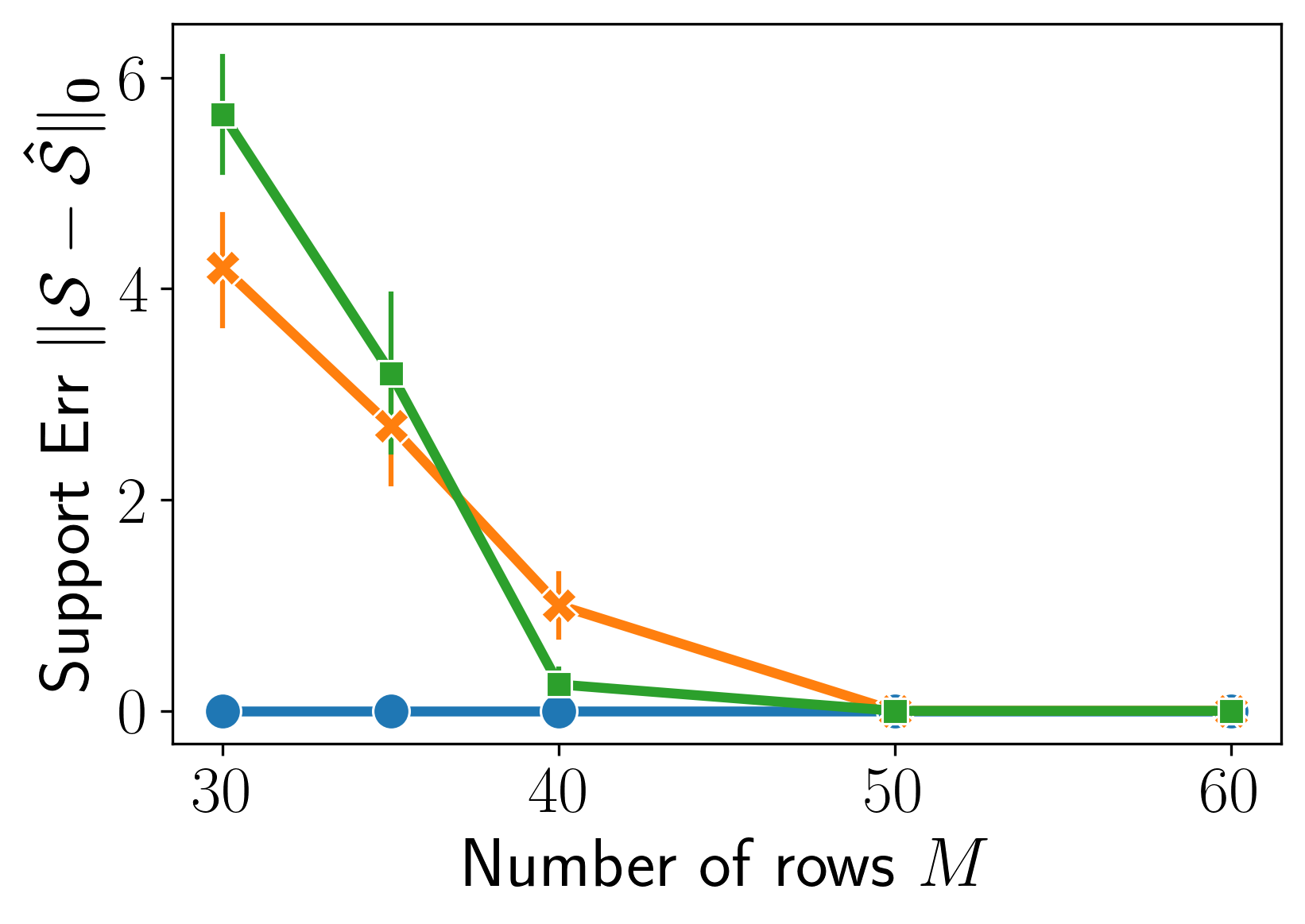}
         \label{fig:fp_supp_vs_m}
     \end{subfigure}
     
        \caption{Results obtained by LightSolver's digital simulator, minimizing the proposed QUBO problem for a 2-bit $\vx$ of length $N=80$. We fixed the cardinality to $k=10$ and set the noise level to $\sigma=0.1$. Error metrics are evaluated over $20$ independent realizations of $\mA,\vx$ and $\vv$. Other details are as in Figure~\ref{fig:motivation}.}
        \label{fig:2-bit}
\end{figure}

\paragraph{2-bit sparse vectors.} We now turn to expand the above experiment, by considering a more flexible representation of $\vx$ with $P=2$.
Recall that our QUBO formulation for the 2-bit setting requires $2N$ spins in total, in contrast to the binary case that requires only $N$ spins. Therefore, we consider a smaller matrix $\mA$ with $N=80$ columns, keeping the total number of spins identical across the two experiments. The 2-bit representation of $\vx$ follows  \eqref{eq:fp_rep}, with $c_i^{\text{min}}=0$ and $d_i=1$; consequently, each \emph{non-zero} entry is randomly sampled from the discrete set $\{1,2,3\}$. Figure~\ref{fig:2-bit} displays the reconstruction and support recovery errors as a function of the sample-size $M$, while fixing the cardinality ($k=10$) and noise ($\sigma=0.1$) levels. As portrayed, the annealer achieves impressive estimation errors, outperforming OMP and lasso especially in the most challenging setting in which the sample size is relatively small.

\section{Conclusions and future directions}\label{sec:conclusions}

In this work, we formulate the ubiquitous sparse coding task as a  quadratic unconstrained binary optimization (QUBO) problem, which, potentially, can be minimized efficiently using quantum computers and Ising machines. Numerical experiments demonstrate the superiority of our quantum-based estimations compared to widely-used, classic sparse approximation algorithms---lasso and OMP. In particular, we report significant improvements in challenging regimes where (i) the sample-size is relatively small, (ii) the noise-level is relatively large, and (iii) the cardinality is relatively large. While our QUBO formulation can handle the most general fixed-point representation of the underlying sparse vector, we believe it would be of great interest to extend the proposed framework to a floating point representation. Such a formulation may reduce the number of spins required to express the solution with a sufficient precision, which is of paramount importance given the challenge of quantum/Ising platforms to increase the number of spins.  We also believe it would be exciting to expand our work beyond regression settings, and provide QUBO formulations to sparse-regularized classification algorithms.

In a broader view, there is a growing interest in using quantum technologies in biological sciences~\cite{emani2021quantum}, and the methods presented in this paper may provide performance improvements in applications where sparse linear regression plays a role. As an example, in genome-wide association studies (GWAS) scientists are interested in accurately identifying which of the thousands of single-nucleotide polymorphisms (SNPs) are associated with high cholesterol, or any other phenotype of interest~\cite{usai2009lasso,waldmann2013evaluation}. The ability to find a sparse, interpretable set of `important' SNPs can be used to improve medical treatment as well as to expand our understanding of the human genome. In these applications, it is common for the number of observations (e.g., individuals participating in a study) to be smaller than the number of features (SNPs), as biotech development allows a massive collection of genetic variants for each individual~\cite{sesia2021false}. The desire to form an interpretable predictive rule together with the high-dimensional nature of the data---the number of SNPs (columns of $A$) is larger than the observations (rows of $A$)---may explain the popularity of sparse linear regression methods in this field. Although conducted on simulated data, the experiments presented in this paper indicate that our quantum-based approach thrives in ultra high-dimensional regimes, and therefore we believe it would be exciting to explore the potential use of our quantum sparse coding method in this domain.


\bibliographystyle{unsrt}
\bibliography{biblio}
\medskip

\appendix

\section{Generating a rectangular matrix with low mutual coherence}
\label{app:genA}

Algorithm~\ref{alg:genA} presents a heuristic iterative procedure for generating a matrix $\mA$ whose mutual coherence $\mu(\mA)$ is minimized. The idea behind this procedure is to design an $\mA$ such that the gram matrix $\mA^{\top}\mA \in \mathbb{R}^{N \times N}$ will be as close as possible to the identity matrix.

\begin{algorithm}
\caption{Heuristic gradient decent algorithm for generating $\mA$ with low mutual coherence.}\label{alg:genA}
\begin{algorithmic}
\Require Number of rows $M$, number of columns $N$, step size $\alpha$, number of iterations $K$.
\Ensure A matrix $\mA$ with normalized columns such that $\mu(\mA)$ defined in \eqref{eq:coherence} is minimized.

\State Initialize $\mA_{1}$ with entries drawn from a Standard Normal distribution.
\State Normalize $\mA_{1}$ such each column $a_i$, $1 \leq i \leq N$, has a unit $L_2$ norm.
\For{$k=1, \dots, K$}
\State $\mA_{k+1} \gets A_k - \alpha \cdot \nabla_{A_k} \| A_k^{\top}A_k - I_{N \times N}\|_F^2$, where $I_{N \times N}$ is the $N \times N$ identity matrix.
\State Normalize $\mA_{k+1}$ to have columns with a unit $L_2$ norm.
\EndFor
\end{algorithmic}
\end{algorithm}

\end{document}